\documentclass[12pt]{article}
\usepackage{multicol}
\usepackage{graphicx}

\usepackage{epsfig}

\begin{document}

%\fancyhead[co]{\footnotesize B. Desplanques et al: Pion charge form factor 
%and constraints from space-time translations}

%\footnotetext[0]{Received November 2009}

\title{Pion charge form factor and constraints \\ from space-time translations}

\author{%
      Bertrand Desplanques$^{1}$\thanks{{\it E-mail address:}  desplanq@lpsc.in2p3.fr},%
\quad Yubing Dong$^{2,3}$\thanks{{\it E-mail address:} dongyb@ihep.ac.cn}\\%
$^{1}$~Laboratoire de Physique Subatomique et Cosmologie, \\
Universit\'e Joseph Fourier Grenoble 1, CNRS/IN2P3, INPG, France\\
$^{2}$~Institute of High Energy Physics, Chinese Academy of Sciences,\\
Beijing 100049, P. R. China\\
$^{3}$~Theoretical Physics Center for Science Facilities (TPCSF), CAS,\\
Beijing 100049, P. R. China\\
}
\maketitle

%\address{%
%1~(Laboratoire de Physique Subatomique et Cosmologie, Universit\'e Joseph
%	Fourier Grenoble 1, CNRS/IN2P3, INPG, France)\\
%2~(Institute of High Energy Physics, Chinese Academy of Sciences,
%Beijing 100049, P. R. China)\\
%3~(Theoretical Physics Center for Science Facilities (TPCSF), CAS,
%Beijing 100049, P. R. China)&&
%}

\begin{abstract}
The role of  Poincar\'e covariant space-time translations 
is investigated in the case of a relativistic quantum mechanics approach 
to the pion charge form factor. 
It is shown that the related constraints are generally inconsistent with the
assumption of a single-particle current, which is most often referred to. 
The only exception is the front-form approach with $q^+=0$. 
How  accounting for the related constraints, as well as restoring 
the equivalence of different RQM approaches in estimating form factors, 
is discussed. Some extensions of this work and, in particular, the relationship
with a dispersion-relation approach, are presented. 
Conclusions relative to the underlying dynamics are given.
\end{abstract}

%\begin{keyword}
%\end{keyword}

%%\begin{pacs}
%\end{pacs}

\begin{multicols}{2}

\section{Introduction}
It is a usual claim that the study of form factors of hadronic systems 
should {\it a priori} provide information on the underlying dynamics. 
Examination of estimates in the framework of relativistic quantum mecha\-nics 
(RQM) nevertheless shows a strong sensitivity to the choice of the form 
used to implement relativity \cite{Dirac:1949cp}, especially 
for the pion charge form factor \cite{Amghar:2004,He:2004ba}. 
The dependence on the form and its associated construction 
of the Poincar\'e algebra in instant, front and point ones
\cite{Bakamjian:1953kh,Keister:sb} results from an incomplete calculation. 
It is expected that it should 
disappear by accounting for many-particle terms in the current 
besides a single-particle one that is usually retained \cite{Sokolov:1978}. 
In absence of the many-particle terms, one can only hope 
that one approach is better than other ones but its choice may reflect 
some own prejudice, necessarily subjective.

Recently, an objective argument, based on pro\-perties of currents 
under Poincar\'e space-time translations, was presented, allowing one 
to discriminate between different  
RQM implementations \cite{Desplanques:2004sp,Desplanques:2008fg}. Invariance 
under space-time translations implies the expected energy-momentum 
conservation but this only represents a part of properties 
that can be ascribed to these transformations in the RQM framework. 
As shown by Lev \cite{Lev:1993}, relations involving the commutator 
of the space-time translation operator, $P^{\mu}$, 
with the currents, which stem from Poincar\'e covariance, have also 
to be verified. Their consideration \cite{Desplanques:2004sp,Desplanques:2008fg} 
suggested a way to account indirectly for the above many-particle currents 
for a scalar system consisting 
of scalar constituents \cite{Wick:1954,Cutkosky:1954}. 
The procedure tends to restore the equality of the squared momentum transferred 
to the constituents and to the whole system, which is {\it a priori} violated 
in incomplete RQM approaches but holds in field-theory ones. 
This work is extended here to a physical system, the pion, 
which represents one of the simplest hadron and, moreover, 
has been the object of extensive studies (see \cite{Desplanques:2009}
for references).

The plan of the paper is as follows. We first review properties 
that currents should fulfill under space-time translations 
and pay a particular attention to constraints they imply 
and go beyond the energy-momentum conservation. 
In the following part, we show how we account for these constraints. 
Results for the  pion charge form factor 
in different forms are then presented together with the result obtained 
when the above constraints are accounted for. 
In the last part, we make various comments in relation with possible extensions.
A conclusion summarizes the main results and the consequences 
that a comparison with experiment suggests for the solution 
of the mass operator used in our calculations.

\section{Covariant space-time translations: new insight and constraints}
Properties  under Poincar\'e covariant space-time translations
imply the following transformations of currents:
\begin{eqnarray}
&&e^{iP \cdot a} \; J^{\nu} (x) \;(S(x)) \; e^{-iP \cdot a}\nonumber \\
&&\hspace*{8mm}=J^{\nu} (x+a) \;(S(x+a)) ,
\hspace*{8mm} \label{eq:1}
\end{eqnarray}
where $P^{\mu}$ represents the 4-momentum opera\-tor. The quantities 
$J^{\nu}(x)$ and $S(x)$ respectively refer to  4-vector and scalar currents.  
When the matrix element of the above relations for $a=-x$ is taken 
between eigenstates of $P^{\mu}$, one obtains:
\begin{eqnarray}
&&\hspace*{-6mm}<i\;| J^{\nu} (x) \;({\rm or}\;S(x))|\;f>
\nonumber \\
&&\hspace*{-4mm}=e^{i(P_i-P_f) \cdot x}<i|J^{\nu}(0) \,
({\rm or} \,(S(0))|f>  .
\hspace*{4mm}\label{eq:2}
\end{eqnarray}
Together with the function $e^{iq \cdot x}$ describing the interaction 
with an external probe carrying momentum $q^{\mu}$, and assuming 
space-time translation invariance, one obtains the standard energy-momentum
conservation relation:
\begin{eqnarray}
(P_f-P_i)^{\mu} = q^{\mu}\, .
\label{eq:3}
\end{eqnarray}
In calculating the matrix element of  Eq.~(\ref{eq:2}), and probably 
for simplicity, it is generally assumed that $J(0)^{\mu}\;({\rm or}\;S(0))$ 
is described by a single-particle ope\-rator. Until recently however, 
it was not checked whether this assumption is consistent 
with further constraints that stem from Eq.~(\ref{eq:1}) 
and were proposed by Lev \cite{Lev:1993}. 
Particular relations of relevance here are the following ones:
\begin{eqnarray}
&&\hspace*{-4mm}\Big[P_{\mu}\;,\Big[ P^{\mu}\;,\; J^{\nu}(x)\Big]\Big]=
-\partial_{\mu}\,\partial^{\mu}\,J^{\nu}(x),
\;\;\; \nonumber \\
&&\hspace*{-4mm}\Big[P_{\mu}\;,\Big[ P^{\mu}\;,\; S(x)\Big]\Big]=
-\partial_{\mu}\,\partial^{\mu}\,S(x) \, . 
\label{eq:4}
\end{eqnarray}
After factorizing the $x$ dependence as in Eq.~(\ref{eq:2}), 
and taking the matrix element of the current, assuming temporarily 
it is a single-particle one, one should get the following relations:
\begin{eqnarray}
&&\hspace*{-6mm}<\;|q^2\; J^{\nu}(0) \;({\rm or}\;S(0))|\;>\nonumber \\
&&\hspace*{0mm}=
<\;|(p_i-p_f)^2\,J^{\nu}(0)\;({\rm or}\;S(0))|\;> \,,\hspace*{6mm} 
\label{eq:5}
\end{eqnarray}
where $q^2$ represents the squared momentum transferred to the system 
and $(p_i-p_f)^2$ the one transferred to the constituents. 
Chec\-king the relations, it is found that they are violated in all cases 
with one exception: the front-form approach 
with the momentum confi\-gu\-ration $q^+=0$. The violation of the relations 
therefore shows that the assumption of a single-particle current is not valid 
in the correspon\-ding approaches, indicating that the current should then 
contain many-particle terms. One can hope that accoun\-ting 
for their contributions would restore the equivalence 
of diffe\-rent approaches in calculating form factors \cite{Sokolov:1978}. 
However, calculating the contribution of many-particle terms 
is quite tedious and this has been done only in a limited number 
of cases \cite{Desplanques:2003nk,Desplanques:2009}. 
Moreover, if they have the effect of restoring the equi\-valence 
with other approaches, they should occur at all orders in the interaction, 
which is apparently hopeless.

\section{Implementation of the constraints}
In order to account for the extra contributions, we observe that the current 
should in one way or another keep the structure of a single-particle one 
as far as this is the case for the front form with $q^+=0$, which fulfills the
constraints. 
Moreover, comparing diffe\-rent approaches, we notice that their expressions 
differ in the coefficient multiplying the momentum transfer $q^{\mu}$ 
and that the diffe\-rence implies interaction effects 
that are here or there depending on the approach. 
These observations suggest to modify the factor of the momentum transfer 
by a coefficient $\alpha$, so that to fulfill Eq.~(\ref{eq:5}). 
We thus obtain the equation:
\begin{eqnarray}
&&\hspace*{-10mm}q^2\!=\!
``[(P_i\!-\!P_f)^2
+2\, (\Delta_i \!-\! \Delta_f)\,  (P_i\!-\!P_f) \cdot \xi
\nonumber \\
&&\hspace*{2cm}+(\Delta_i \!-\!\Delta_f)^2 \;\xi^2]"
\nonumber \\
&&\hspace*{-6mm}=\alpha^2q^2-2\,\alpha ``(\Delta_i \!-\!\Delta_f)" \;q \cdot \xi
 \nonumber \\
&&\hspace*{2cm}+``(\Delta_i \!-\!\Delta_f)^2"\;\xi^2 \,,
%\nonumber \\ 
\label{eq:6}
\end{eqnarray}
where $\xi^{\mu}$ represents the orientation of the hyperplane 
on which physics is described and $\Delta$ holds for an interaction effect. 
It is immediately seen that, for the front-form case with $q^+=0$, 
the above equation is sa\-tis\-fied with $\alpha=1$  
as $\xi^2=0$ and $\xi.q=0$. In this case, the equality 
of the squared momentum transferred to the system 
and to the constituents, Eq. (\ref{eq:5}),  is tri\-vially fulfilled. 
In the other cases, one has to take into account the modification 
of the calculation given by the coefficient $\alpha$, 
which is solution of Eq. (\ref{eq:6}).

\section{Some results}

The detail of the implementation of the constraints  stemming from
space-time translations has been given in Ref.~\cite{Desplanquesun:2009} 
for a scalar system  like the pion one consisting of two spin-1/2 particles. 
Recovering the equivalence of different approaches may require 
a particular current but, in the case of the charge form factor, 
the current so obtained is often the one that is expected, 
probably because it fulfills minimal requirements 
such as the existence of an underlying conserved current 
or the invariance of the charge (unaffected by the constraints) under boosts. 
For the wave function, we use the one obtained  
in an independent work \cite{Desplanques:2009}
aimed to study the asymptotic behavior of the form factor in RQM approaches. 
The corresponding mass operator involves both a confinement 
and an instantaneous one-gluon exchange interaction.

Results are presented in Fig. 1 for various  approaches 
(see Refs.~\cite{Desplanques:2004sp,Desplanques:2008fg} for definitions 
and further details). 
The left panel, which involves low $Q^2$, is sensitive 
to the squared charge radius while the right panel covers 
the remaining range of $Q^2$ where data are available.
\end{multicols} 
\vspace{-0mm}
\centerline{\rule{150mm}{0.1pt}}
\vspace{0mm}
%\ruleup
%
\begin{figure}[htb]
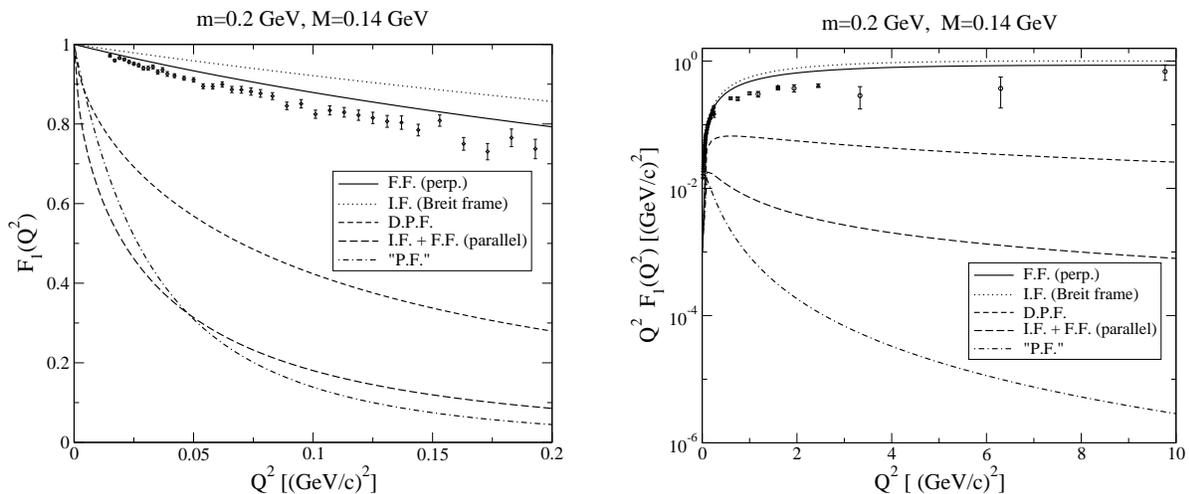

%\ruleup
%\mbox{ \epsfig{ file=pi1s.eps, width=7.4cm}}
%\hspace*{1cm}
%\mbox{ \epsfig{ file=pi1S.eps, width=7.4cm}}
\mbox{ \epsfig{ file=pi1s.eps, width=7.4cm}
\hspace*{0.6cm}
 \epsfig{ file=pi1ss.eps, width=7.4cm}}
\caption { \label{fig1} Pion charge form factor in different forms 
and effect of the constraints related to the covariance properties 
of the current under space-time translations: left panel 
for the low-$Q^2$ range and right panel for the intermediate-$Q^2$ range 
(see text for references about further details and some definitions). 
The dot, short-dash, long-dash, dot-dash and continuous curves represent 
results without the effect of constraints discussed here. 
When this effect is included, all curves become
identical to the front-form one with $q^{+}=0$, which is unchanged 
(continuous curve denoted F.F. (perp.)).}
%\ruledown
\vspace{-4mm}
\end{figure} 
%\ruledown
\vspace{-0mm}
\centerline{\rule{150mm}{0.1pt}}
\vspace{0mm}
\begin{multicols}{2}
Examination of the figure shows a considerable discrepancy 
between different approaches, which is very similar 
to the spinless-constituent 
case \cite{Amghar:2002jx,Desplanques:2004sp,Desplanques:2008fg}. 
At low $Q^2$, the discrepancy exhibits in some cases a dependence 
of the squared radius on the inverse of the squared pion mass. 
It is noticed that the Lorentz invariance of the form factor obtained in an
earlier point-form approach (``P.F.")\cite{Bakamjian:1961} 
or in  a point-form approach inspired from Dirac's one 
(D.P.F.) \cite{Desplanques:2004rd} does not guarantee a better result. 
When the effect of the implementation of the constraints related 
to space-time translations is accounted for, all form factors 
become identical to the front-form one with $q^+=0$, 
which was fulfilling the constraints from the start.
Hence, the full restoration of properties related to the Poincar\'e covariance
of these transformations is essential to obtain reliable results.

\section{Further developments}
While showing numerically that expressions of the pion form factor 
in  different forms could give the same results after accounting 
for constraints related to space-time translations, we wondered 
what could be the common expression behind this important property. 
Following different works in the scalar constituent or in the spin-1/2 cases
\cite{Melikhov:2001zv,Krutov:2001gu,Desplanques:2008fg}, 
it is found that this expression could be identified to the one based 
on a s-channel dispersion-relation  approach, which is explicitly 
Lorentz invariant and, moreover, fulfills constraints related 
to space-time translations. This expression, which is not well known, reads:
\begin{eqnarray}
&&\hspace*{-0.8cm}F_1(Q^2) =\frac{1}{N} \!\int\! d\bar{s} \;  d(\frac{s_i\!-\!s_f}{Q}) \; 
 \phi(s_i) \; \phi(s_f)
 \nonumber \\
&& \hspace*{2.0cm}\times\frac{ 2\sqrt{s_i\,s_f}
\;\theta(\cdots) }{D\sqrt{D}} \, ,
\label{eq:disp}
\end{eqnarray}
where the variables and various quantities are defined 
in Ref.~\cite{Desplanques:2008fg}.
Obtaining this result in each form supposes that the correspon\-ding
implementation of re\-la\-tivity represented by the Bakamjian-Thomas 
construction of the Poincar\'e algebra in the instant form \cite
{Bakamjian:1953kh} and its generalizations in the other
cases \cite{Keister:sb} has been consistently performed. 
It also supposes a non-trivial change  of variables. 
We notice that the above expression confirms the one given 
in Ref.~\cite{Melikhov:2001zv} but disagrees 
with the one given in Ref.~\cite{Krutov:2001gu}. 
For this last work, the discre\-pancy factor, $(s_i+s_f+Q^2)/(2\sqrt{s_i\,s_f})$, 
is the same as for the scalar-constituent case \cite{Desplanques:2008fg}.

We mentioned that accounting for constraints related 
to space-time translations was amounting to implicitly consider 
the contribution of many-particles currents. 
These currents only represent a minimal subset, 
which is required to restore some symmetry properties. 
In the present case of the pion charge form factor, 
they do not allow one to reproduce its asymptotic expression. 
This one supposes to consider specific two-particle currents  
within the RQM approach \cite{Desplanques:2009}.

It is not rare that the breaking of some symmetry can provide unexpected
results, at the limit of paradoxes. The variation of the charge radius with the
inverse of the pion mass obtained in some cases is one of them (more binding
produces a larger radius!). As shown here, accounting for constraints from
Poincar\'e covariant translations corrects for this surprising result.

\section{Conclusion}
We have considered the role of constraints relative 
to space-time translations on the estimate of the pion charge form factor. 
Accounting for these constraints amounts to restore 
the equality of the squared momentum tranferred to the constituents 
and to the pion, which is violated in the simplest RQM calculation 
but is trivially fulfilled in field theory. 
As these constraints stem from a covariant transformation 
of currents under space-time translations, accounting for them 
represents a necessary ingredient of a covariant calculation of form factors.
A complementary insight is the following one. In RQM approaches, changing the
underlying hypersurface for another one implies interaction effects. One is not
therefore surprised that the simplest calculation of elastic form factors
depends on its choice.  Accoun\-ting for the constraints discussed in this work
amounts to consider further interaction effects that remove the
differences implied by this choice, as expected from a fully Poincar\'e
covariant calculation.

Considering the front-form results with $q^+=0$ as representative 
of the results obtained in different approaches after accounting 
for constraints related to space-time translations, 
it appears that the calculated form factor tends to overestimate 
the measured one. As there was no optimisation of the estimate, 
one can think that a better value of the pion decay constant, $f_{\pi}$, 
could help to explain the measurements in the low-$Q^2$  range. 
The statement is based on the relation $r_{\pi}^2=3/(4\pi^2f_{\pi}^2)$ 
and a calculated value of $f_{\pi}$ which overestimates 
the measured one (106  and 93 MeV respectively). 
A better value of $f_{\pi}$ could be obtained by diminishing the quark mass 
or by reducing the weight of high-momentum components 
in the solution of the mass operator that was used. 
This second alternative would have the advantage 
to also reduce the discre\-pancy in the intermediate-$Q^2$ range.

{\bf Acknowledgments} This work is partly supported by the National Sciences Foundations of China
under grants No. 10775148,10975146 (Y.B.). The author is also grateful to the
CAS for grant No KJCX3-SYW-N2. B.D. is very grateful to IHEP for offering 
hospitality after the conference.

\end{multicols}

\vspace{-2mm}
\centerline{\rule{80mm}{0.1pt}}
\vspace{2mm}

\begin{multicols}{2}

\end{multicols}

\clearpage

\end{document}